\def \K{~\rm{K}}

\def \AU{~\rm{AU}}

\def \yr{~\rm{yr}}

\documentclass[12pt,preprint]{emulateapj}
\usepackage{natbib}
\shorttitle{Blue HB Stars}
\shortauthors{Soker \& Harpaz}

\begin{document}

\title{OVERLUMINOUS BLUE HORIZONTAL BRANCH STARS FORMED BY LOW MASS COMPANIONS}

\author{Noam Soker\altaffilmark{1} \& Amos Harpaz\altaffilmark{1}}
\altaffiltext{1}{Department of Physics, Technion$-$Israel
Institute of Technology, Haifa 32000 Israel;
soker@physics.technion.ac.il; phr89ah@aluf-old.technion.ac.il.}

\begin{abstract}
We construct a speculative scenario for rotation-induced extra helium mixing to
the envelope of horizontal branch (HB) stars.
This scenario differs from previous ones in that the mixing occurs after
the star has left the red giant branch (RGB).
We follow the evolution of a low metallicity star from the RGB to the
HB, and examine the density profile and radius in the core-envelope boundary region.
In the transition from the RGB to the HB the envelope shrinks by two
orders of magnitude in size and the core swells, such that any non-negligible rotation
on the RGB will result in a strong rotational shear at the core-envelope boundary.
For a non-negligible rotation to exist on the RGB the star has to be spun up by a
companion spiraling inside its envelope (a common envelope evolution).
We speculate that shear instabilities on the HB
might mix helium-rich core material to the envelope.
The shallow density profile on the HB is less likely to prevent mixing.
As previously shown, extra helium mixing can account for the
overluminous blue HB stars found in some globular clusters.
Although being speculative, this study supports the idea that the presence of
low mass companions, from planets to low mass main sequence stars,
influence the evolution of stars, and can explain some properties of the
color-magnitude (Herzsprung-Russel) diagram of globular clusters.
Namely, low mass companions can be an ingredient in the so called `second parameter'
of globular clusters.

\end{abstract}


\section{INTRODUCTION}
\label{intro}

Evolved sun-like stars which burn helium in their cores occupy the
horizontal branch (HB) in the Hertzsprung-Russel (HR) diagram.
The distributions of stars on the HB (the HB morphology; also
referred to as the color-magnitude diagram [CMD])
differ substantially from one globular cluster (GC) to another.
It has long been known that metallicity is the main, but not sole,
factor which determines the HB morphology
(for a historical review see, e.g., Rood et al. 1997;
Fusi Pecci \& Bellazzini 1997).
 The other factor (or factors) which determines the HB morphology is
termed the `second parameter'.
 For more than 30 years it is clear that mass loss on the red giant branch
(RGB) is closely connected with the second parameter (Rood 1973),
as well as processes that form abundance variations (Gratton et al. 2004).

Sweigart \& Catelan (1998; also Sweigart 1999; Moehler et al. 1999;
Moehler {\it et al.} 2000) claimed that mass loss on the RGB by itself
cannot be the second parameter, and it should be supplied by another
`noncanonical scenario', e.g., rotation or helium mixing, which
requires rotation as well.
Behr {\it et al.} (2000) found the second parameter problem to be
connected with rotation, and noted that single star evolution cannot explain
the observed rotation of HB stars, even when fast core rotations are considered.
Soker \& Harpaz (2000) argued that fast rotating HB stars have been
probably spun-up by planets, brown dwarfs, or low-mass main sequence
stars, while they evolved on the RGB, further supporting the
`sub-stellar second parameter' model (Soker 1998).
In the `sub-stellar second parameter' model the mass loss process on the RGB
is influenced by the interaction of the RGB star with a planet,
a brown dwarf, or a very low mass main sequence companion.
An attractive feature of the binary-planet second parameter model is that it makes
a connection with sdB stars, which are blue HB (BHB) in the field;
many of the the sdB stars have a very close companion (Maxted et al. 2001).
{{{    In the binary-planet (sub stellar) second parameter model the companions are
very light, and will be destroy as they deepen to the envelope
(Soker 1988). Therefore, the non-detection of companions to HB stars in
GCs (Moni Bidin et al. 2006) is not in contradiction with the model. }}}

In a recent paper Catelan et al. (2006;
{{{   see also Rich et al. 1997, Pritzl et al. 2002, and Raimondo et al. 2002) }}}
find that the BHB stars
and RR Lyrae stars in the GC NGC 6388 are overluminous.
Namely, NGC 6388 has the same age and chemical composition as the GC 47 Tucanae,
and the stellar distribution on the CMD is very similar, except for the BHB.
Catelan et al. (2006) argue that a non-canonical second parameter
is required to explain this difference
{{{   (see discussion in Catelan 2007). }}}
More over, only $\sim 17 \%$ of the stars in NGC 6388 should
be influenced by this second parameter, which they suggest can be
enhanced helium abundance (Y) and/or the helium-core mass at He-flash.
Similar results hold for the GC NGC 6441, {{{   but more stars are
influenced by the non-canonical second parameter than in NGC 6388 }}}
(Caloi \& D'Antona 2006).
{{{   Motivated by the recent results of Catelan et al. (2006), and
difficulties with all other non-canonical second-parameter candidates
(Catelan 2007), }}}
we examine whether the planet second parameter model can in principle account for
the differences of the HB morphologies of these clusters
by the fast envelope rotation induced by the companion.
{{{   In a previous paper (Soker 1998) the basic model was proposed, where the companions
lead to high mass loss rate on the RGB, hence bluer stars on the HB.
Indeed NGC 6388 and NGC 6441 have very blue HB. }}}
We propose (speculate) that if the helium core flash occurs in a fast rotating RGB star,
then helium might be mixed to the envelope as the star approaches the HB.

\section{THE BEHAVIOR DURING THE HELIUM CORE FLASH}
\label{flash}

We present results of one-dimensional (spherical) calculation
of a star evolving from the upper RGB to the HB.
The numerical code is based on the code that was used by Harpaz \& Kovetz (1981).
The composition used is $Z=0.0015$ and $X=0.7875$,
and the opacity is taken from Seaton et al. (1994).

We start our description just before the first helium flash; this is our initial model.
At this stage the star has a core mass of $M_c=0.49 M_\odot$, a total mass of
$M_\ast = 0.7 M_\odot$, a radius of $R_\ast=170 R_\odot$, and a luminosity of
$L_\ast = 2100 L_\odot$. The low mass results from mass loss along the RGB.
Most of the envelope is convective, from $m=0.4978$ to $m= 0.6993 M_\odot$; only at
its bottom a mass of $M_{\rm rad}=0.008 M_\odot$ is radiative.

For comparison with most of the overluminous HB stars a lower envelope mass is required,
such that the star be hotter when reaching the HB.
However, stars with lower envelope mass might more easily mix helium during the core flash
{{{    (Brown et al. 2001; }}} Cassisi et al. 2003; Lanz et al. 2004),
so we conservatively took the extreme more problematic case.
In any case, we concentrate on the core-envelope boundary, and the envelope
mass has little influence on the conclusions.

As it is well known (e.g., Dearborn et al. 2006) the first helium flash
does not occur at the center.
More helium flashes occur as the star approaches equilibrium on the zero-age HB
(e.g., Dearborn et al. 2006 who termed them miniflashes).
Our final model is given after the star has relaxed following the final helium miniflash.
The radius and luminosity are $R_\ast = 8 R_\odot$ and $L_\ast = 48 L_\odot$, hence
the effective temperature is $T_{\rm eff}= 5300 \K$.
The transition time from the RGB to the HB is several millions years.

\begin{figure}  
\vskip 0.5 cm
\resizebox{0.4\textwidth}{!}{\includegraphics{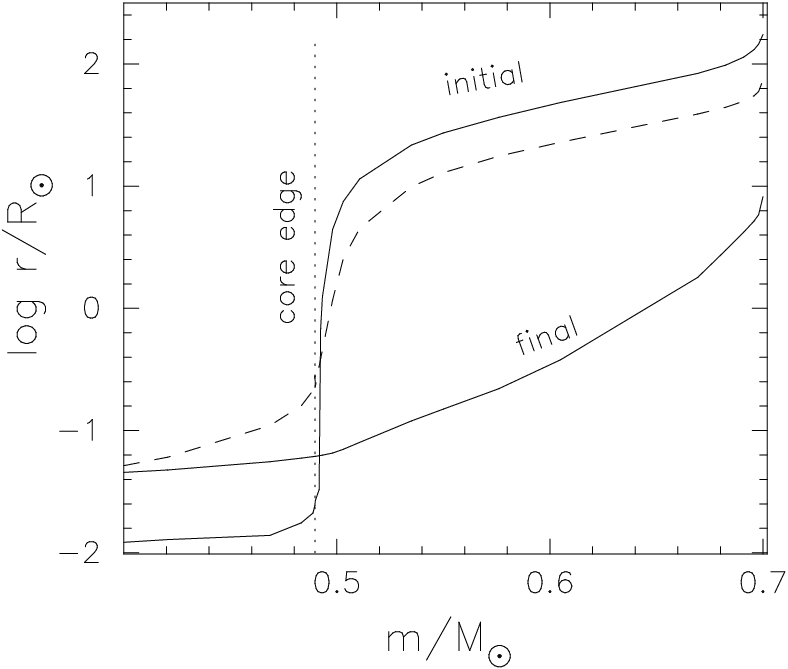}}
\vskip -0.0 cm  
\caption{Radius, in units of solar radius, as function of mass,
in units of solar mass, of the stellar model.
Model are given just before the first helium flash at $t=0$ (marked `initial'),
at the middle of the transition period, at $t=5 \times 10^5 \yr$  (dashed line), and
after the star relaxed following the last helium flash, at $t=3 \times 10^6 \yr$
(marked 'final').
The vertical dotted line marks the edge of the core, where hydrogen mass fraction
is a few per cents. The hydrogen burning shell is about one scale height above this
edge.}
\label{radius}
\end{figure}

\begin{figure}  
\resizebox{0.4\textwidth}{!}{\includegraphics{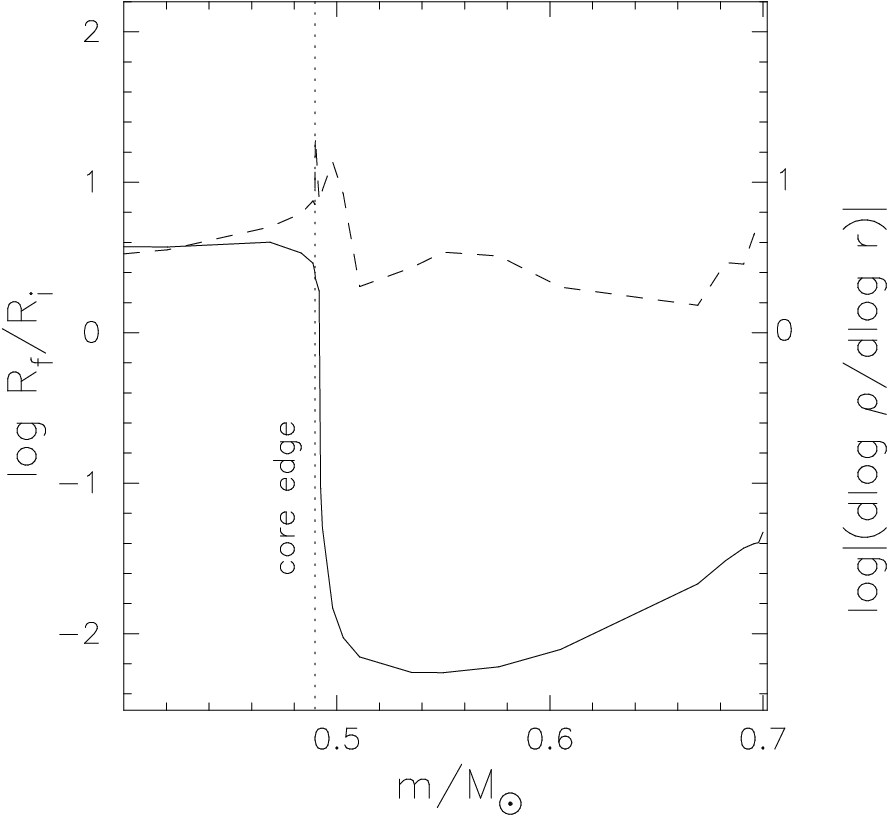}}
\vskip .5 cm  
\caption{Solid line: Ratio of final to initial radius as function of mass
(scale on the left side). Dashed line: Ratio of radius to density scale height
(scale on the right side).
}
\label{ratio}
\end{figure}

In Figure \ref{radius} we present the radius as function of mass in the initial
and final models, as well as at an intermediate stage;
the vertical line marks the boundary of the core,
which for our purposes can be considered as the outer region of the helium-rich
core region. From this figure we see that while the envelope contracts
by a large factor, the core expands by a large factor.
This is further emphasized in Figure \ref{ratio}.
Therefore, conservation of specific angular momentum implies
that the envelope will spin-up while the helium-rich core will slow down.
For a non-negligible initial rotation profile this will lead to a strong
shear at the core-envelope boundary of the newly born HB star.
In Figure \ref{ratio} we also give the ratio of the radius to the density
scale height $r/H_\rho=\vert(d \ln \rho/d \ln r )\vert$.
The pressure scale height is smaller than the density scale height by $\sim 20 \%$.

\begin{figure}  
\vskip 0.1 cm
\resizebox{0.4\textwidth}{!}{\includegraphics{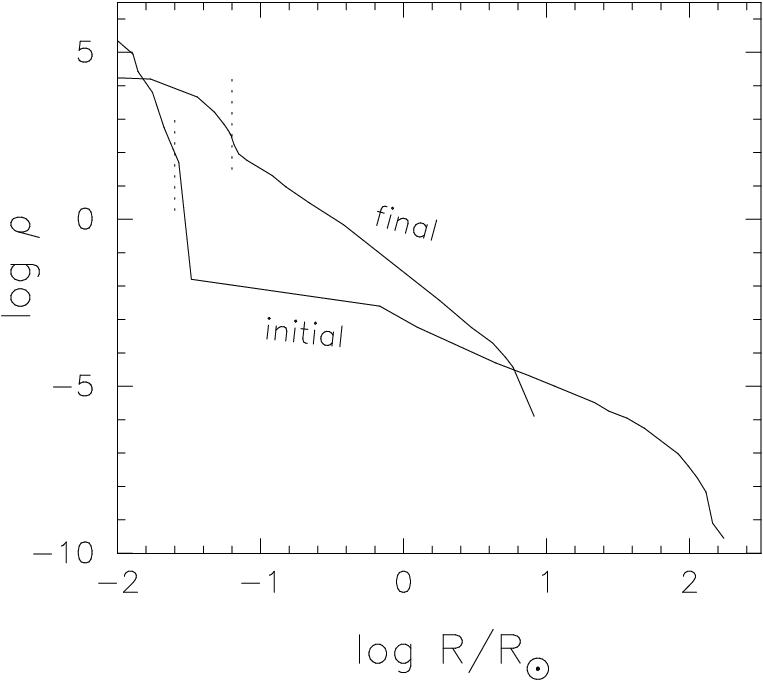}}
\vskip . cm  
\caption{The density profile versus stellar radius just before the helium shell
flashes start and after the star has relaxed on the horizontal branch.
Vertical lines mark the core outer boundary.}
\label{density}
\end{figure}

In Figure \ref{density} we present the initial and final density profiles. From
this figure we learn that on the RGB the star has a huge
density contrast between the core and envelope. This will reduce the
efficiency of mixing between the core and the envelope even if strong shear exists.
On the HB the density contrast has substantially decreased, and it is very modest.



\section{HELIUM MIXING}
\label{mixing}

Extra mixing of core material to the envelope was considered in the past
(e.g., Langer et al. 2000; Denissenkov \& Tout 2000; Suda \& Fujimoto 2006),
{{{    and its relevant to the overluminous HB stars is discussed by
Moehler \& Sweigart (2006a). }}}
The extra mixing occurring along the RGB (Recio-Blanco \& De Laverny 2007,
and references therein) does not mix sufficient helium to the envelope
to account for the over-luminous HB stars.
Basically, helium mixing on the RGB seems not to work efficiently, and
if it were, it would contradict observations (Denissenkov \& Tout 2000).
Siess \& Livio (1999) considered a low mass companion that is destroyed
in the inner radiative zone (between the bottom of the convective envelope
and the hydrogen burning shell) of an RGB star, and studied the response of
the RGB star to the mass added in that region.
At a high companion-mass destruction rate the convective zone reaches down to the
hydrogen burning shell, and can mix nuclear-processed material to the envelope.
However, the dense helium-rich core prevents mixing of large quantities of helium.

Suda \& Fujimoto (2006) considered three modes of mixing of helium from the core to
the envelope of RGB stars. In one of these modes they studied a mixing following
the helium flash. Our proposed mixing mechanism is different in that we consider
the mixing to occur after the star shrinks to the HB.

Extra mixing of helium to the envelope might occur also during the core helium
flash of post-RGB stars {{{    (Brown et al. 2001; }}} Cassisi et al. 2003; Lanz et al. 2004).
However, these delayed (late) flashers form very hot HB stars, much hotter than most
over-luminous HB stars considered by Catelan et al. (2006).
Based on the evolution from the RGB to the HB described in section 2 we
propose that mixing of helium takes place for stars that undergo helium flashes
on the RGB, when there is still a substantial envelope mass.
The mixing occurs as the star approaches the HB and starts its HB phase.
Mixing of other elements from core to envelope can still occur during the RGB,
but we attribute most helium mixing to the early HB phase.
The condition, we propose, is that the RGB star was spun-up by a low mass companion.
Lattanzio et al. (2007) simulated the rotation of the core during the helium flash,
and found no extra mixing. Our proposed scenario is different than their in
that we consider the shear between the envelope and the core,
rather than the core rotation.

Our proposed shear mixing is motivated by earlier studies of rotation-induced mixing in
RGB stars (Denissenkov \& Tout 2000; Suda \& Fujimoto 2006,
and references therein), as well as of studies of
shear mixing in accretion disks around WD (Rosner et al. 2001) and in
rotating asymptotic giant branch (AGB) stars (De Marco \& Soker 2002).
De Marco \& Soker (2002) suggested that the fast spinning of the AGB envelope
layer above the core is caused by a low mass companion spiraling deep inside the
giant envelope.
The destruction of a companion near the core of an RGB star can also lead to the
formation of an accretion disk. However, due to the large density contrast between
the core and the envelope during the RGB (Figure \ref{density}), we consider mixing
of large quantities of helium unlikely during the RGB phase.
Instead, we speculate that extra mixing occurs due to the shear between the expanding core
and the contracting envelope as the star approaches the HB.

The shell laying one density scale hight ($H_\rho$) above the core in the final
model contracts by a factor of $\sim 100$ from its initial radius.
If it conserve angular momentum the ratio of its angular speed to that
of the Keplerian angular speed, $\Omega_f$, increases by a factor of $\sim 10$.
Using equation (3) of Suda \& Fujimoto (2006; or eq. 20 of Denissenkov \& Tout 2000)
for the criterion for sear
instability, with the pressure scale height at the core boundary and the
molecular weight difference between core and envelope, we find that for this
final velocity to cause an instability it should be $\Omega_f \ga 0.3$.
Therefore, its initial value should be $\Omega_f \ga 0.03$.
Such a rotation velocity on the RGB requires a binary companion to
spin-up the envelope.
The common envelope phase should not occur at early RGB stages.
If it does occur at early stages of the RGB, then the RGB star loses
too much mass until it experience the helium core flash.
Such a mass loss will substantially slow down the RGB star (Soker \& Harpaz 2000).

The shear will mix helium-rich material from the core to the envelope layers
just above the core.
Meridional circularization will spread the helium in the envelope on the Eddington-Sweet
time scale (Tassoul 2000)
$t_{ES}=\Omega_f^{-2} G M_\ast^2 / L_\ast R_\ast= 4 \times 10^5 (\Omega_f/0.3)^{-2} \yr$,
where in the second equality we have substituted the values for the final model.
This is much shorter than the HB life time of the star.
In our scenario the fast rotation induced by a companion is required both for
mixing helium from the core to the envelope and for spreading the helium in
the envelope.
We note that mixing by meridional circularization was proposed before, but
on the RGB (e.g., Sweigart \& Mengel 1979; Denissenkov \& Tout 2000).

{{{   Caloi \& D'Antona (2006) find that the required enrichment in their model of
helium pollution at star formation epoch is up to $Y \sim 0.4$.
Let us assume that such an enrichment is required in the envelope of
the HB stars in the present scenario. }}}
For an envelope mass of $\la 0.1 M_\odot$ an extra of $\la 0.01 M_\odot$
of helium is needed to be mixed.
Such a mass exists from the core's edge down to about
a density scale height ($\sim H_\rho$ deep).
The extra mass that is mixed is of the order of the mass in the envelope's inner
radiative region in the initial model $M_{\rm rad} = 0.008 M_\odot$.
Namely, even if only the radiative inner region on the RGB kept its fast rotation
from the RGB to the HB, and not the entire convective RGB envelope (on the HB
{{{    a large fraction of the envelope, or even the entire envelope for hot HB stars, }}}
is radiative), the required helium-rich mass can be
mixed before the region above the core slows down much.

{{{ One observational limitation on the model is that carbon-rich material will not
be mixed to the envelope. Large amounts of carbon are formed during the
helium flashes.
Paczynski \& Tremaine (1977; also Thomas 1967)
suggested that helium flashes occurring far from the core center can
lead to carbon mixing to the envelope, and the formation of carbon stars.
Such stars are not observed in GCs.
In our model the outer $\sim 0.01M\odot$ of the core is basically carbon-free.
But a mass of up to $\sim 0.02 M_\odot$ is needed to be mixed
to the envelope (Sweigart, A., private communication 2007).
Deep core regions contain $\sim 5 \%$ of carbon by mass.
However, our model is likely to overestimate the mixing of carbon to the
very outer core region ($\sim 0.02-0.03 M_\odot$).
First, mixing may take time and will not be completed during the flashes
(Lattanzio et al. 2007). Our model assumes instantaneous mixing in the convective region.
Second, the outer region of the core may rotate quite fast (before it expands) as it
is made of recently burned envelope material, and in our model the envelope has
a relatively large angular momentum. Using the Solberg-Hoiland criterion for stability
may leave the outer region of the core stable to convection.
We only need that the outer convective region moves one scale height deeper.
The question whether too much carbon is mixed to the envelope in our model should be
addressed with more sophisticated models.  }}}

\section{DISCUSSION AND SUMMARY}
\label{summary}
Using a one-dimensional stellar evolution code we followed the
transition of a low metallicity star from the RGB to the HB (Section 2).
The goal was to emphasize the change in the
properties of the core-envelope boundary region. While on the RGB
the outer core region, which is helium-rich, is $\sim 4-5$
orders of magnitude denser than the envelope inner region.
On the zero-age HB, on the other hand, the density profile is much
shallower (Figure\ref{density}).
The much shallower density profile is less likely to prevent mixing of
helium-rich core material to the envelope as a result of shear instabilities.
Namely, over one or two pressure (or density) scale heights near the core-envelope
boundary, where the mixing takes place, much more mass is involved in HB stars
than in RGB stars.
In the RGB to the HB transition the core swells while the envelope shrinks
by an order of magnitude or more (Figure \ref{radius}).
As a result, the core spins down and the envelope spins up.
Even a modest rotation of the envelope on the RGB will result in a strong
shear in the newly formed core-envelope boundary region.

We speculate (Section 3) that shear instabilities might mix helium-rich
gas from the core to the envelope layers laying just above the core, after the
star reaches the HB.
We further suggest that meridional circularization will then spread the helium
in the entire envelope and increase its helium fraction up to $Y \sim 0.4$.
Such a high helium fraction lead to brighter HB stars (Sweigart \& Gross 1976;
Sweigart 1987), and
can account for the overluminous blue HB stars found in the GCs NGC 6388
(Catelan et al. 2006) and NGC 6441 (Caloi \& D'Antona 2006).
The lower core mass, because of the mixing of $\la 0.01 M_\odot$ of helium to
the envelope, reduces somewhat the HB stellar luminosity, but the effect
is much smaller than that due to the enhanced helium abundance in the envelope
(Sweigart et al. 1987).

Caloi \& D'Antona  (2006) proposed that the high helium abundance in NGC 6441
is a result of self-enrichment by star formation episodes in the early
evolution of the GC. This scenario
{{{    was supported by Moehler \& Sweigart (2006a), }}} but it encounters two
difficulties.
First, the high helium abundance for $\ga 14 \%$ of all stars will hold also on
the main sequence and RGB if primordial helium enrichment occurred.
Catelan et al (2006), find no difference between the main sequence and
RGB of NGC 6388 and 47 Tuc; these two GCs are different only in their HB.
{{{   However, the observations of NGC 6388 done so far are not sensitive
enough to find evidence for helium enrichment in the main sequence.
But NGC 2808 does show a blue main sequence, containing $\sim ~15\%$ of
the stars, which apparently corresponds to $Y\sim 0.4$ (D'Antona et al. 2005). }}}
Seconds, as Caloi \& D'Antona  (2006) noted, AGB stars cannot account for
helium abundance of $Y>0.35$ required for $\sim 14 \%$ of the HB stars in NGC 6441.
{{{   We admit that the helium self-enrichment of GC at star formation
(Caloi \& D'Antona  2006), termed primordial contamination,
better fits observations than our proposed binary model (see discussion in
Catelan 2007). However, because of
some difficulties with this model as well (Catelan 2007), we conducted our research
to examine whether the binary model can also account for the
tilted HB morphology (overluminous HB stars). }}}

We attribute fast rotation to the effect of a low mass companion (Soker \& Harpaz 2000)
depositing orbital angular momentum to the envelope on the RGB via a common envelope
evolution.
The companion can survive the evolution, or be destructed near the core.
Sills \& Pinsonneault (2000) studies the possibility of explaining fast rotating
HB stars by single star evolution.
For reasons listed in Soker \& Harpaz (2000) single stars
cannot account for fast rotation on the RGB and of HB stars;
for more discussion and different views, see Recio-Blanco et la. (2002).
For the present purpose, there is another problem with fast rotating
single star scenario.
Fast rotation not only influence the HB, but also increase the turnoff
(from the main sequence) age (Mengel \& Gross 1976).
This is not observe when comparing NGC 6388 with 47 Tuc (Catelan et al. 2006).

{{{   The model has two predictions that can be put to test in the near future.
\newline
(1) The mixing mechanism requires fast rotation in the core-envelope boundary.
This does not imply directly that the HB star itself be fast rotator.
In most cases of RGB stars swallowing planets the RGB wind will carry a large fraction of
the angular momentum that was deposited to the envelope, and will slow down the envelope.
Bluer HB stars have much lighter envelopes, and therefore are expected to be slower rotators
(Soker \& Harpaz 2000).
What is needed for the present model is that the radiative zone of the RGB star
will maintain its fast rotation.
Therefore, if the envelope mass is very low at the helium core flash, the mass loss
during the transition to the HB is sufficient to slow the convective envelope down.
The inner radiative region of the envelope will not have time to lose angular momentum.
However, in many cases the mass loss will not be enough to slow down the envelope, and
we expect the HB star to rotate, such that its angular momentum cannot be explained
by a single star evolution. In other cases the radiative zone will have time to
lose its angular momentum, and no helium mixing will occur.
Moehler \& Sweigart (2006b) found no indication for rotation in three overluminous stars
in NGC 6388. If many more overluminous stars are found with no indication of rotation
then our proposed mechanism is in trouble.
Either the mixing occurs already on the RGB and the star has time to lose
most of its angular momentum (Soker \& Harpaz 2000), or the convective envelope
slows down during the transition to the HB but not the inner radiative
region. But more likely the binary model altogether is not the explanation for the
overluminous HB star.

We do note that in the present model, and in any other model where the helium mixing
occurs  during or after the transition to the HB, it is expected that higher
temperature HB stars be more overluminous, as found in NGC 6388 and NGC 6441
(Catelan \& Sweigart 2006; Caloi \& D'Antona 2006).
The reason is that the helium mass that is mixed to the envelope
is fixed by the amount of rotation, and it is limited by the depth of less than
about one density scale height. Bluer HB stars have less envelope mass, and they will
contain higher fraction of helium therefore.

(2) The proposed model requires the presence of a low mass companion close to the envelope.
We predict that planets in GCs do exist, but at orbital separations
of $0.3 \AU \la a \la 3 \AU$. This range is larger
than that of the planets found in the solar neighborhood.
Weldrake et al. (2007) looked for transits in the GCs $\omega$ Cen and 47 Tuc,
and did not find any.
The null detection of planets in 47 Tuc {{{   (Gilliland et al. 2000) }}}
is not a problem to the binary model as this GC
has only few stars on its blue HB.
The reason why no planets are expected in GCs with a small population of blue HB stars
was explained by Soker \& Hadar (2001).
Basically, close planets, or brown dwarfs, or low mass main sequence stars,
will be swallowed by the RGB star and enhance its
mass loss rate, hence leading to the formation of BHB stars.
If the planets enter the envelope early on the RGB, the RGB star will slow down
substantially via its mass loss process (Soker \& Harpaz 2000).
Therefore, if most planets (or brown dwarfs or low mass main sequence stars)
are close to the their parent star, the model predicts BHB stars even
in metal-rich clusters, but no helium mixing will occur as the RGB stars will
be slow rotators by the time they experience the core helium flash.
In clusters where planets form at larger orbital separations, and they
are massive enough to spiral all the way to the RGB core, we expect helium
mixing to occur when the star reaches the HB.

The condition that the companion will reach the inner radiative region
of the envelope as it spirals in during the common envelope phase,
gives the lower mass range of the companion.
A low mass companion will be evaporated when the escape speed from its
surface is about equal to the sound speed in the RGB envelope
(Soker 1998).
In the RGB model the radiative region is inner to $\sim 1-3 R_\odot$
(depending on the exact envelope mass and luminosity).
Using equation (1) from Soker (1998), we find the condition to be
$M_2 \ga 0.01 M_\odot$.
The upper limit on the secondary mass is given by the requirement that the
companion will not expel the entire envelope before it comes close to the core.
This depends on the envelope mass when the common envelope phase starts,
but can crudly put as $M_2 \la 0.2 M_\odot$.
Over all, the condition on the companion mass is
\begin{equation}
0.01 M_\odot \la M_2 \la 0.2 M_\odot.
\label{mass2}
\end{equation}

Soker \& Hadar (2001) studied the correlations between the HB morphology
and some other properties of GCs, and found that some correlations, e.g., of HB
morphology with the central density and core radius, exist only for GCs within a
narrow metallicity range (see also Recio-Blanco et al. 2006).
Soker \& Hadar (2001)  conjectured that the lack of correlations with {\it present}
properties of GCs (besides metallicity), is because the variation
of the HB morphologies between GCs having similar metallicities
is caused by a process, or processes, whose effect was determined
at the {\it formation time} of GCs.
They then argued that the `sub-stellar second parameter' model fits
this conjecture.
This is because the processes which determine the presence
of low mass companions and their properties occur during the formation
epoch of the star and its circumstellar disk.

In any case, we do predict that close planets, brown dwarfs and/or very low mass main sequence
stars, will be eventually found around many of the main sequence stars in non-canonical
second parameter GCs, but at larger orbital separations than known planets around
stars in the field.
The statistics of Weldrake et al. (2007) is low, but if more searchers of GCs with
large BHB population find no such companions up to $\sim 1 AU$,
then the binary second parameter model will have to be abandon.
}}}

Despite the speculative nature of the presently proposed extra helium mixing,
we view this paper as a significant contribution to the
`sub-stellar second parameter' model (see section 1), because it suggests
a way by which the model can account for extra helium mixing.

{{{    I thank M\'{a}rcio Catelan,  Allen Sweigart, Sabine Moehler ,
Alejandra Recio-Blanco, Santi Cassisi, John Lattanzio, and Francesca D'Antona
for helpful comments. }}}
This research was supported by a grant from the Asher Space Research
Institute at the Technion.

\end{document}